\documentstyle[prl,aps,epsfig,twocolumn]{revtex}
\begin{document}
\title{Mixing of magnetic and phononic excitations in 
incommensurate Spin-Peierls systems}
\author{Ariel Dobry}
\address{Departamento de F\'{\i}sica, Universidad Nacional de Rosario, \\
and Instituto de F\'{\i}sica Rosario, Avenida Pellegrini 250,\\
2000 Rosario, Argentina}
\author{David Ibaceta}
\address{Instituto de Astronom\'{\i}a y F\'{\i}sica del Espacio,\\
Casilla de Correos 67, Sucursal 28,\\
1428 Buenos Aires, Argentina.}
\maketitle

\begin{abstract}
We analyze the excitation spectra of a spin-phonon coupled chain in the
presence of a soliton. This is taken as a microscopic model of a
Spin-Peierls material placed in a high magnetic field. We show, by using a
semiclassical approximation in the bosonized representation of the spins
that a trapped magnetic state obtained in the adiabatic approximation is
destroyed by dynamical phonons. Low energy states are phonons trapped by the
soliton. When the magnetic gap is smaller than the phonon frequencies the
only low energy state is a mixed magneto-phonon state with the energy of the
gap. We emphasize that our results are relevant for the Raman spectra of
the inorganic Spin-Peierls material CuGeO$_3$.
\end{abstract}


The discovery in 1993 by Hase et al \cite{Hase} of the first inorganic
Spin-Peierls compound CuGeO$_3$ has opened the possibility of study the
physics of this collective phenomena in a deep way. Several
experimental proves have given an exhaustive information about the
excitation spectra of this system and its evolution with an applied magnetic
field. The effect of non-magnetic impurities has been investigated also.

Theoretical studies have focused on simplified magnetic model. The excitation
spectra in the low temperature phase have been analyzed using a dimerized
and frustrated Heisenberg chain as a minimal model for this material\cite
{UhrigS,RieraK,Bouzerar}. The logic underlying these studies are: the
competition between magnetic and elastic energies resolves in the low
temperature phase in the dimerization of the lattice. Once this process
takes place, phononic and magnetic excitation completely decouple 
and the magnetic
excitations are the same as the chemical dimerized system. This point of
view is based on an adiabatic approximation supposing that the energy scale
of the magnetic process are high enough respect to the phononic ones. As it has
been recently emphasized \cite{Uhrig}, this relation is not fulfilled for
CuGeO$_3$ where the phonons relevant for the dimerization process are about
one order of magnitude more energetic than the magnetic gap. The adiabatic
approximation is questionable for this system. An antiadiabatic approach has
been developed. The frustrated interaction arises, in this context, from the
integration of the in-chain phonons and the explicit dimerization from the
interchain interaction treated in a mean field approximation\cite{Uhrig2}.
The same frustrated-dimerized Hamiltonian is therefore obtained but with a
reinterpretation of the parameters. What is clearly missed in these studies is
a general understanding on how spin and phonons mix as elementary excitation
and how the spectra of Spin-Peierls systems is built as a result of this
mixing. Some recent numerical results have partially addressed this question 
\cite{Augier}.

In this paper, we analyze the excitation spectra of one-dimensional
spin-phonon system by semiclassical techniques on the bosonized representation
of the spins subsystem. We focus on the properties of this system in a high
magnetic field. In the dimerized phase the system is in a singlet ground
state. Coupling with a magnetic field is not effective up to a critical
field value where some dimer break and their spins are freely from their 
singlet. The lattice also relaxes in this process forming the so called
soliton lattice. This relaxation is a first indication that spin and phonons
do not act as independent excitation. It has been experimentally shown by RX
measurements that the lattice becomes incommensurate 
following a soliton pattern\cite{Kiry}. The magnetic profile of the soliton 
has been analyzed by NMR measurement \cite{fagot}.

More recently, optical proves have shown the spectral signature of the
incommensurate phase. In the uniform dimerized phase a low energy resonance
appears at 30 cm$^{-1}$ which is a smaller energy than two times the magnon
gap\cite{Lemmens}. This peak has been adjudicated to a bound state of two
triplet \cite{Bouzerar}. When the incommensurate phase appears the spectral
weight is transfered from this state to a lower energy peak at the position of
the magnon gap\cite{Loa}. Direct magnon process could not be seen in the
optical response due to spin conservation. Soliton assisted one magnon
excitation was recently been proposed as the origin of this peak\cite{Loa}
in similarity with the situation in the presence of non-magnetic impurities%
\cite{Els}.

This approach is based on the image of the soliton as an isolated spin in
the externally dimerized chain so that an adiabatic approximation is
supposed. We will show that this state in fact disappears when the dynamics
of the phonons is included . We will also show that a trapped magneto-phonon
state with the energy of the gap appears in the antiadabatic regime so
explaining the optical data.

Let us proceed more formally. Our starting Hamiltonian for an Spin-Peierls
compound is:

\begin{eqnarray}
H=H_{ph}+H_{mg}  \label{H}
\end{eqnarray}
\begin{eqnarray}
H_{ph}=\sum_i\frac{P_i^2}{2M}+\frac K2(u_{i+1}-u_i)^2
\end{eqnarray}
\begin{eqnarray}
H_{mg}=\sum_iJ(1+\frac \alpha 2(u_{i+1}-u_i)){\bf S}_i\cdot {\bf S}_{i+1}
\end{eqnarray}
${\bf S}_i$ are spin-1/2 operators of the $i$ ion and $\alpha $ the
magneto-elastic coupling . $H_{ph}$ represents our simplified model for the
phonons. It contains a scalar coordinate $u_i$ and its conjugate momentum $%
P_i$ which are supposed to be the relevant ion coordinates for the
dimerization process. As we will only retain the phonons relevant for SP
transition, the specific dispersion of the phonons is not important in our
approach. The tridimensional character of the phonons field is essential to
account for the finite Spin-Peierls temperature and  the excitations in the
low temperature dimerized phase\cite{domain}. We will discuss later  its
effect on the incommensurate phase.

The low energy spectrum could be studied by bosonization. The spin variables
are approximately represented by the bosonic field $\phi (x)$ and its
conjugated momentum $\Pi (x)$\cite{Affleck}. In addition, we retain only the
phonon modes producing a smooth deviation of the dimerized pattern. So, we
make the replacement $(-1)^iu_i\rightarrow u(x)$. The low energy Hamiltonian
becomes: 
\begin{eqnarray}
H_{ph} &=&\int dx\{\frac{aP(x)^2}{2M}+\frac{2K}au(x)^2\}  \nonumber \\
H_{bos} &=&\int dx\{\frac 1{2\pi }[\frac{v_s}\eta (\partial _x\phi (x))^2 
\nonumber \\
&&+v_s\eta (\pi \Pi (x))^2]+\frac{J\alpha u(x)}{\pi a_0}\sin \phi (x)\}
\label{Hbos}
\end{eqnarray}
where $a$ is the lattice constant of the original chain and $a_0$ a short
range cutoff introduced in the bosonization procedure. $v_s$ is the spin
wave velocity and $\eta $ is related with the exponent of the correlation
functions. For the isotropic Heisenberg model with nn interaction we have $%
v_s=\frac{Ja\pi }2$ and $\eta =2$.

We use a semiclassical approach in the form of a self consistent harmonic
approximation(SHA) as it has been originally proposed by Nakano and Fukuyama (NF) 
\cite{NF1}. In this approach the boson field $\phi $ is split in a classical
component $\phi _{cl}$ and the quantum fluctuation $\hat{\phi}$. The last
term of (\ref{Hbos}) is developed up to second order around $\phi _{cl}$ and
then treated self-consistently by the following replacement:

\begin{eqnarray}
\sin\phi \rightarrow  \nonumber \\
e^{-\langle \hat{\phi}^2\rangle /2}[\sin(\phi_{cl})(1-\frac{(\hat{\phi}^2-
\langle \hat{\phi}^2\rangle)}2)+\cos(\phi_{cl})\hat{\phi}]  \label{SHA}
\end{eqnarray}
$\langle \hat{\phi}^2\rangle $ is the ground state expectation value.

In their original work NF have fixed the displacement field $u$ to its
equilibrium classical value ($u_{cl}$), so that an adiabatic approximation
was assumed. Let us summarize the main results of this study and its
consequence for the spectra:

\begin{itemize}
\item  The classical equations for $\phi $ and $u(x)$ have a soliton
solution of the form: $\sin (\phi _{cl})=\tanh (x/\xi )\nonumber\\ %
u_{cl}(x)=u_0\tanh (x/\xi )$ . $\xi $, the soliton width, is $v_s/\Delta $
and $\Delta $ is the gap over the homogeneous dimerized state (the 'magnon'
gap) and $u_0$ are the displacements in this homogeneous dimerized state.
The soliton carries $S_z=1/2$ spin. In the presence of an external magnetic
field these solitons will condense in the ground state and accommodate in a
soliton lattice structure.

\item  The creation energy of the soliton as well as the excitations over
this solitonic vacuum could be evaluated once the eigenvalues of the
fluctuation operator are known. This eigenvalue problem corresponds to a one
of a Schroedinger-like equation which reads as:

\begin{eqnarray}
\{-\nabla _{\hat{x}}^2+\tanh ^2(\hat{x})\}\psi _\lambda =\frac{\Omega
(\lambda )^2}{\Delta ^2}\psi _\lambda (\hat{x})  \label{sch} \\
\hat{x}\equiv \frac x\xi   \nonumber
\end{eqnarray}
$\Omega (\lambda )$ are the frequency oscillations in presence of the
soliton. Eq. (\ref{sch}) has one bound state at $\frac{\Omega _b}{\Delta}
=\sqrt{\frac{\sqrt{5}-1}2}\sim 0.786$ and a continuum started at $\Omega =\Delta $. The
excitation spectra of the theory is spanned by the following state:
\end{itemize}

\begin{itemize}
\item  {The ground state of the quantum soliton builded around $\phi _{cl}$.
Their creation energy ($E_s$) is given by its classical energy plus the
difference between the sum of the zero point energies of the oscillators $%
(\Omega (\lambda ))$ and the ones in absence of the soliton. This creation
energy measures the critical field for the commensurate- incommensurate
transition\cite{NF1}.}

\item  {The excited state of the soliton with energy $E_s^{*}=E_s+\Omega _b$.
 It is related with the bound state of Eq. (\ref{sch}).}

\item  {Labeling by q the continuum of level of Eq. \ref{sch}, they are $%
\Omega (q)=v_s\sqrt{q^2+1/\xi ^2}$ This is just the kinetic energy of a
particle in the soliton free sector. In terms of the original spin chain
this is the low energy dispersion of the $S_z=0$ component of a magnon.
Therefore this state corresponds to the scattering of a magnon in presence
of the soliton. Moreover, when one of the continuum modes is excited once,
we get a two particle magnon-soliton state.}
\end{itemize}

The two last adjudication of states have not been done in the original
analysis of the NF work. 
As our formalism breaks SU(2) symmetry at intermediate stage, the total spin
of each state is not directly accessible and should be carefully reanalyzed.
We have previously stated that the states below the continuum correspond to
the emission of a magnon in the presence of the soliton. Therefore they have
total spin-3/2. The bound state has the same total spin as the soliton. 
Then, it is an optical active mode and it is the analogue in our formalism
of the so called soliton assisted magnon process founded in the strong
dimerization limit\cite{Els}.

Moreover, the previous predicted spectra could be tested by numerical
 exact diagonalization on finite chain. To this end, we solve iteratively
the adiabatic equations for $u_i$ arising from Hamiltonian Eq. \ref{H}. 
The ground state of the spin system was recalculated at each iteration step
by Lanczos diagonalization. 
We considered chains of odd number of site up to $L=23$ sites with periodic 
boundary conditions 
and look for the equilibrium positions in the subspace $S_z=1/2$.
The numerical details of the method was given in \cite{Feiguin,Sch}.

\begin{figure}[htb]
\epsfig{file=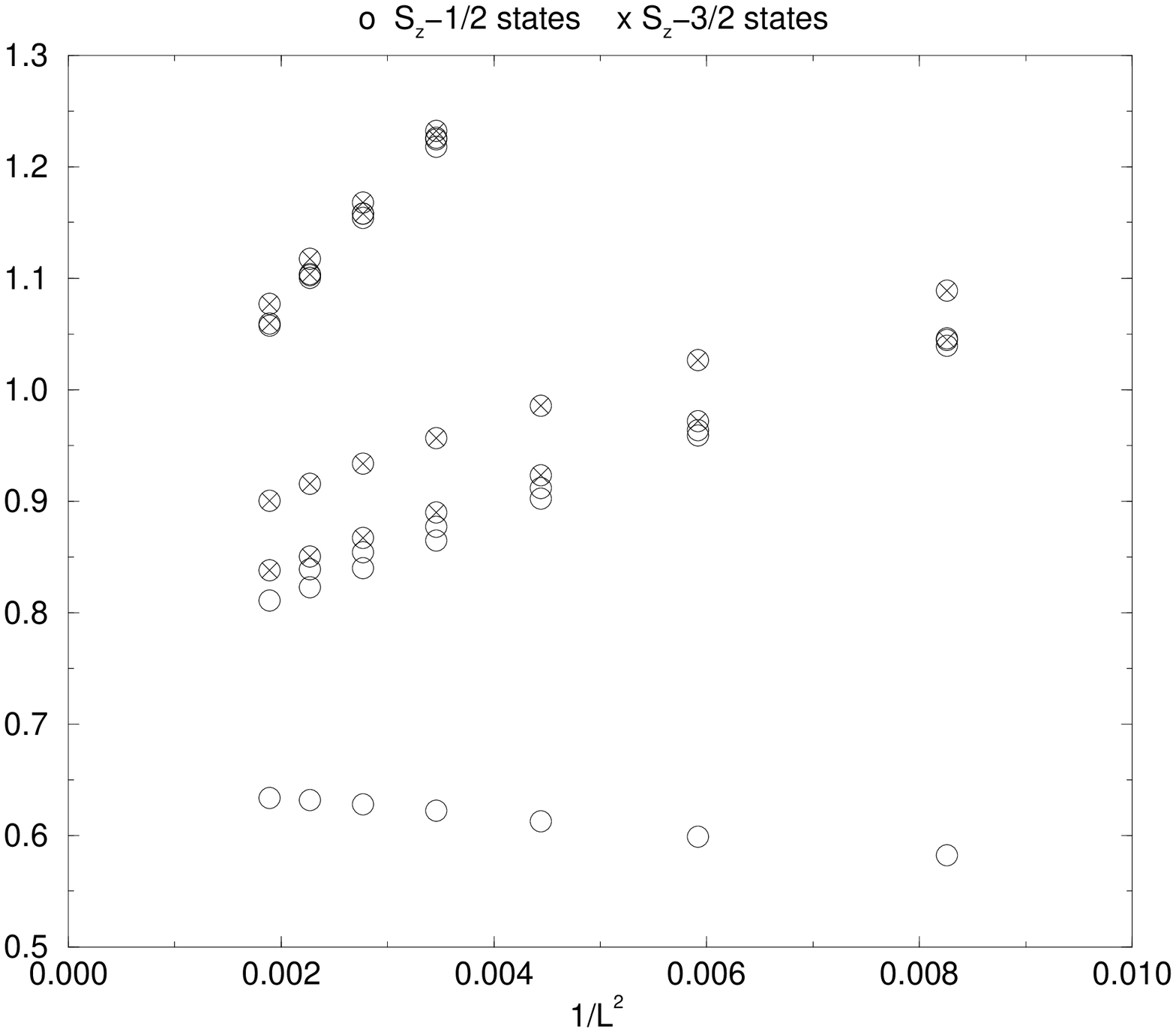,width=6cm,angle=-90}
\vskip .5 truecm
\caption{ The excitation spectrum calculated by Lanczos diagonalization
in the presence of an adiabatic soliton.
Circles are for states of $S_z=\frac12$ and crosses for $S_z=\frac32$.
The zero of the energies is chosen at the soliton ground state.
Energies are in unit of $J$.}
\label{fig1}
\end{figure}

Once the ionic coordinate converge 
we diagonalize the spin Hamiltonian, with fixed ${u_i}$,
 to obtain the low lying
excitation in the $S_z=1/2$ and $S_z=3/2$ subspaces. The results are shown in
Fig.\ref{fig1} as a function of the $1/L^2$.  
Parameter $\frac{K}{(\alpha J)^2}=\frac14$ has been chosen in order
to have a thin soliton and to reduce finite size effect.  
The bound state predicted by the bosonized theory is clearly seen.
The higher energy states collapse in a continuum of total spin-3/2
in the thermodynamical limit. By linear extrapolation, we found 
for $L\rightarrow \infty$ the ratio $0.84$ between the bound state 
and the border
of the continuum. This value compares well with our previous prediction
$0.786$. We conclude that the method gives at least a qualitative
features of the low energy spectra.
The  states appearing in the border of the continuum in Fig.
 (\ref{fig1}) are an artifact of the strongly thin soliton
 we are considering.
We have check that for bigger values of $\frac{K}{(\alpha J)^2}=\frac14$
these states in fact disappear.

We discuss in the following  two effects that could change our estimated value
of the bound state from the SHA method:

\begin{enumerate}

\item{ In going from \ref{H} to \ref{Hbos}, we have neglected as NF in their
original work a term
proportional to $cos(2\phi)$ in the bosonized theory. 
Even though this operator plays an important role in the case of
Ising anisotropy, it is a marginally irrelevant operator in the 
Heisenberg case and therefore it should not change  the qualitative features
previously discussed. 
 Moreover, NF have studied in a further contribution
\cite{NF2} the effect of this term. Unfortunately, the estimation
of the bound state value is strongly dependent on the value
of the cuttoff in the bosonization procedure ($a_0$) in this situation.
If we assume as NF that $\pi a_0=a$, the bound state becomes  
${\Omega _b}{\Delta}=\sqrt{\frac{\sqrt{3}}2}\sim 0.93$.
We can conclude as a general fact that the presence of this term tend
to increase the bound state value.}

\item{It has been recently noted \cite{Fabrizio,Uhlig3}
that a full self-consistent treatment (where 
$ <\phi^2>$ in Eq. \ref{SHA} is not fixed to its value 
in the ground state) changes the form of the soliton.
Moreover, it has been shown in Ref.(\cite{Uhlig3}) that this
treatment produces a smoother soliton than the one considered here.
Therefore, we conclude that this treatment will reduce the
 bound state value.}

\end{enumerate}
   
In the following we analyze the evolution of the spectra when
non-adiabatic effects are included. For simplicity we do not introduce
the two additional effects previously discussed. These effects will
not affect the prediction of the trapped magneto-phonon state we will
find.

{\em Inclusion of non-adiabatic effects.} To go beyond this static adiabatic
approximation in this semi-classical calculation it is necessary to include
fluctuations in the displacement field $u(x)$. Therefore, we split $u(x)$ as $%
u_{cl}+\hat{u}$ and replace it in Eq.(\ref{Hbos}). The classical equations are
the same as before. The fluctuation operator is now a 2 x 2 differential
operator with component over $u$ and $\phi$. The eigenvalue problem is now
given by:

\begin{equation}
\begin{array}{lllll}
\Delta^2[-\nabla^2_{\hat{x}} + \tanh^2(\hat{x})]\psi_1 & + & \frac{u_0}{a}%
8\pi K v_s {\rm sech}(\hat{x})\psi_2 & = & \Omega^2 \psi_1 \\ 
\omega^2 u_0{\rm sech}(\hat{x}) \psi_1 & + & \omega^2 \psi_2 & = & \Omega^2
\psi_2
\end{array}
\label{fluc2}
\end{equation}
$\psi_1$ and $\psi_2$ are the component of the fluctuation eigenvector over $%
\phi$ and $u$ respectively. $\omega=\sqrt{\frac{4K}{M}}$ is the phonon
frequency at $q=\pi$ The eigenfrequency are as previously given by $\Omega$.
We obtain from the second equation: 
\begin{eqnarray}
\psi_2=\frac{u_0 \omega^2{\rm {sech}(\hat{x})}} {\Omega^2-\omega^2}\psi_1
\label{psi12}
\end{eqnarray}
Replacing in the first equation we have a kind of Schroedinger equation
where the potential depends on the energy. This equation reads:

\begin{eqnarray}
\{-\nabla^2_{\hat{x}} + \tanh^2(\hat{x})\}\psi_1+ (\frac{\omega}{\Delta})^2 
\frac{{\rm {sech}}^2(\hat{x})}{(\frac{\Omega}{\Delta})^2- (\frac{\omega}{\Delta})^2}%
\psi_1=(\frac{\Omega}{\Delta})^2\psi_1  \label{schef}
\end{eqnarray}
It is worthy to note that,
 the only relevant parameter is $\frac{\omega}{\Delta}$.Then, we can
follow the evolution of the spectra from the adiabatic to the antiadiabatic
limit by moving this parameter.

\begin{figure}[htb]
\epsfig{file=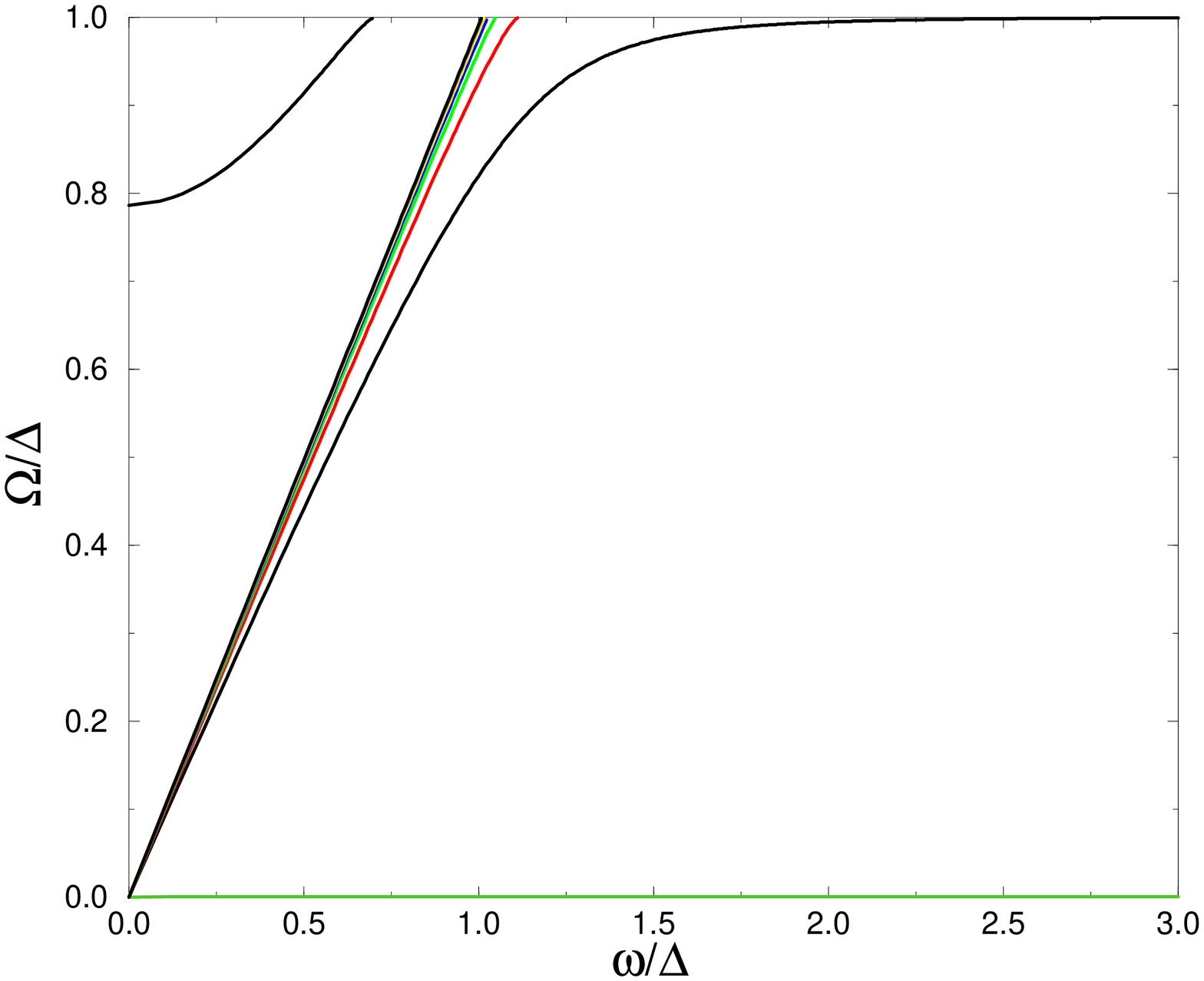,width=6cm,angle=-90}

\vskip .5 truecm

\caption{ The excitation spectrum as a function of the phonon frequency.
Both energy scales are given in units of the magnon gap ($\Delta$)}
\label{fig2}
\end{figure}

We have numerically solve Eq. (\ref{schef}). In Fig. \ref{fig2}, we show the
evolution of the excitation spectra as a function of $\frac \omega \Delta $.
The soliton creation energy has been taken as the zero of the energy.

The main results are:
\begin{itemize}
\item  {There is a zero energy solution which was not present in the
adiabatic calculation. This mode arises in the complete translation
invariance of the theory. It is nothing but the translational mode of the
soliton. In a realistic situation solitons are not isolated, they form a
soliton lattice. This zero mode will acquire some dispersion giving rise to
the so called phason modes of the soliton lattice. If the magnetic field is
not much higher than the critical one the solitons will be rather separate
and we can neglect their interference. }

In the direction perpendicular to the magnetic chains, solitons will
accommodate in a domain wall, i.e. they form an array of parallel solitons.
Their coherency is assured by the interchain elastic coupling. The zero mode
as well as all the states of the spectra will have a transversal dispersion,
they will depend on the momentum perpendicular to the chains. The spectrum
shown in Fig. (\ref{fig2}) gives the q=0 excitations. We will analyze their
dispersive character in a forthcoming work.

\item  {The upper branch corresponds to the excited state of the soliton
previously found in the adiabatic approximation. Its energy increases with $%
\omega $ up to a critical value $\frac{\omega ^2}{\Delta ^2}=\frac 12$ where
this state disapear. For this $\omega $ and $\Omega =\Delta $, 
the energy dependent
potential vanishes. The lattice soliton produces an harmonic potential over the
magnetic soliton where it oscillates. In the adiabatic approximation the
lattice soliton is frozen and the magnetic one oscillates. For finite $%
\omega $, the lattice soliton can also move and vibrate with respect to the
magnetic deformation. Finally, at the critical value of $\omega $ there is
no more a localized vibration and the soliton becomes an unique entity both
for the lattice and for the spins. This explains the disparition of the
trapped state at a critical phonon frequency.}

\item  {The straight line of slope one in Fig.\ref{fig2} corresponds to $%
\Omega =\omega $. The eigenstate with this energy corresponds to the
excitation of a phonon in the presence of the soliton. The eigenvalues near
above are satellite phonons corresponding to phonons trapped by the soliton
(i.e. particle vibrations near the soliton). They have only spectral
weight on the phonons as it can be seen putting in Eq.(\ref{psi12})
 $\Omega \sim \omega $.
For increasing phonon frequency, most of these eigenvalues lose but the
lowest survives acquiring the energy of the magnetic gap. This state is a
mixing of a magnetic and a phononic excitation as we show in the following. }

\item  {Four Peierls active phonons have been identified for CuGeO$_3$\cite
{wernergross}. They have frequencies of 3.12, 6.53, 11.1 and 24.6 THz i.e.
the smaller one is about 150K. The spin gap is rather small for this system $%
\Delta =24K$. As we previously stated the real parameter regime where this
material lives is $\omega >\Delta $ and the only survival bound state has
the energy of the gap. This is a singlet state which we associate with the
low energy peak seen in Raman scattering in the incommensurate phase \cite
{Loa}. The $\psi _1$ and $\psi _2$ components of the corresponding
eingenfunction are different from zero for this state.}
\end{itemize}

Their mixed character could be advocated by analyzing separately the
magnetic and phononic spectral response. The magnetic Raman operator of a
dimerized chain is given by ${\cal R}_{mg}=\sum_i(1+\gamma _{mg}(-1)^i{\bf S}%
_i\cdot {\bf S}_{i+1})$\cite{Muthukumar} $\gamma _{mg}$ is a microscopic
parameter. In our bosonized formalism the most relevant contribution comes
from the staggered part and it is given by $\frac{\gamma _{mg}}{\pi a_0}%
sin(\phi )$. By using the SHA given by Eq. (\ref{SHA}), retaining the term
linear in $\hat{\phi}$ and developing in the basis of the eigenvalues of the
fluctuation operator the magnetic spectral weight ($M_n$) of an state of
energy $\Omega _n$ is given by:

%

\begin{eqnarray}
M_n = \frac{\gamma_{mg}^2}{2 \Omega_n}\left\{\frac{4 K}{J \alpha a} \frac{%
\Omega_n- \omega^2}{\omega^2} \int \psi_2 dx \right\}^2
\end{eqnarray}
where we have used Eq. (\ref{psi12}) to write $\psi_1$ as a 
function of $\psi_2$.
Phonons contribution to Raman scattering is proportional to the square of
the transition elements of the normal coordinates ${\cal R}_{ph}=\gamma
_{ph}\int \frac{dx}au(x)$, the phonons spectral weight of the n-state is
given by:

\begin{eqnarray}
P_n = \frac{\gamma_{ph}^2}{2 \Omega_n}\left\{ \int \psi_2 \frac{dx}a
\right\}^2
\end{eqnarray}

The relation between the magnetic and phononic contribution to the n-th
peak of the spectra is:

\begin{eqnarray}
\frac{M_n}{P_n} = \frac{\gamma_{mg}^2}{\gamma_{ph}^2} \left\{ \frac{4 K}{%
J\alpha}\right\}^2 \left\{ \frac{\Omega_n-\omega^2}{\omega^2} \right\}^2
\label{MP}
\end{eqnarray}

We take the first two factors as a measure of the spin-phonon coupling , we
fix this parameters when moving $\omega $. Eq. (\ref{MP}) shows that the
magnetic component of a given Raman peak increases when its position shifts
from a phonon frequency. The lower energy eigenvalue of Fig({\ref{fig2})
starts being a trapped phonons for $\frac \omega \Delta <1$ and transmutes in
a magneto-phonon in the opposite limit where the real material lives. Its 
existence is a consequence of the quasi-onedimensional character of our
system because a one-dimensional well has always one bound state. }
 
In summary, we have shown that nonadiabatic effects are relevant for
incommensurate Spin-Peierls system. In the antiadiadatic limit applicable to
CuGeO$_3$ we show the apparition of a trapped mixed state with the energy of
the magnon gap. It gives an alternative explanation of the Raman induced peak
found in the incommensurate phase of CuGeO$_3$.
A.D. acknowledges G. Uhrig  and A. Greco for useful discussion.
We are grateful to J. Riera for useful discussions and computational help. 
We are grateful to P. Lemmens for discussions and for
pointing out Ref. \cite{Loa}.

\end{document}